\documentclass[twocolumn,amsmath,amssymb,aps]{revtex4}
\newcommand{\Vg}{$V_\text{G}$}
\newcommand{\mueV}{$\upmu$eV}
\usepackage{graphicx}
\usepackage{dcolumn}
\usepackage{bm}
\usepackage{upgreek}
\begin{document}
\preprint{APS/123-QED}

\title{Locating environmental charge impurities with confluent laser spectroscopy of multiple quantum dots}
\author{M. Hauck$^{1}$, F. Seilmeier$^{1}$, S. E. Beavan$^{1}$, A. Badolato$^{2}$, P. M. Petroff$^{3}$, 
and A. H\"ogele$^{1}$}

\affiliation{$^1$Fakult\"at f\"ur Physik, Munich Quantum Center and Center for
NanoScience (CeNS), Ludwig-Maximilians-Universit\"at M\"unchen,
80539 M\"unchen, Germany}

\affiliation{$^2$Department of Physics and Astronomy, University
of Rochester, Rochester, New York 14627, USA}

\affiliation{$^3$Materials Department, University of California,
Santa Barbara, California 93106, USA}
%

\date{\today}

\begin{abstract}
We used resonant laser spectroscopy of multiple confocal InGaAs
quantum dots to spatially locate charge fluctuators in the
surrounding semiconductor matrix. By mapping out the resonance
condition between a narrow-band laser and the neutral exciton
transitions of individual dots in a field effect device, we
identified spectral discontinuities as arising from charging and
discharging events that take place within the volume adjacent to the quantum dots.
Our analysis suggests that residual carbon dopants are a
major source of charge-fluctuating traps in quantum dot
heterostructures.
\end{abstract}

\maketitle

\section{Introduction}

The exciton transitions in self-assembled InGaAs quantum dots
(QDs) are elementary to potential applications in quantum
information processing \cite{Imamoglu1999} and quantum
cryptography \cite{Shields2007}. For quantum cryptography
protocols, QDs can be used to generate indistinguishable single
photons \cite{Santori2002, Gazzano2013} with high repetition rates
\cite{Michler2000}, or to produce entangled photon pairs on demand
\cite{Stevenson2006}.  In addition, efficient all-optical spin
manipulation schemes characteristic to QDs \cite{Atature2006} can
be exploited for spintronics applications \cite{Warburton2013}.
Recent developments in spin-photon
interfacing can also be used to reversibly transfer qubits between
light and QD states \cite{Gao2013,DeGreve2012} and place QDs
alongside the nitrogen-vacancy center in diamond \cite{Pfaff2014}
as a potential solid-state building block for practical quantum
devices. All these experiments ubiquitously rely on a well defined
and stable resonance condition between the exciton transition and
the laser fields.

In current QD devices, however, the fidelity of such protocols is
limited by spectral fluctuations. Early resonant experiments
identified spectral diffusion as a primary limitation to the
temporal stability of the resonance condition \cite{Hogele2004a}.
More recent studies of resonance fluorescence \cite{Muller2007,
Vamivakas2009} and its dynamics \cite{Kuhlmann2013} found
that the main source of resonance instability is the charge noise
due to fluctuations in the electrostatic environment, which is
detrimental to the quality of single photons that can be generated
in QD devices \cite{Nguyen2011, Matthiesen2012, Prechtel2013}.
Recent work on related device heterostructures has identified
charge traps at the GaAs/AlGaAs superlattice (SL) interface as a
major source of spectral diffusion \cite{Houel2012a}, and similar
effects have also been observed in devices without a SL
\cite{Nguyen2013}. In this work, we investigate the resonance
condition for a number of QDs in a field-effect device, and find
that spectral jumps are caused by charge fluctuations occurring in
the semiconductor volume surrounding the QDs, and are not purely
an interface effect. Using the gate-voltage dependence combined
with the magnitude of the spectral fluctuations, we identify the
likely source of these charge traps as residual carbon impurities,
and the individual impurity sites can be located more precisely
when their influence can be observed in more than one QD spectrum.
Such spectroscopic studies could be used in the first instance as
a highly sensitive measure of semiconductor purity, and secondly
to adjust the growth methods and heterostructure design so as to
reduce the detrimental charge-noise in QD devices.

\section{Experimental Details}

The self-assembled InGaAs QDs studied here were grown by molecular
beam epitaxy (MBE) \cite{Leonard1993} with subsequent annealing,
and have emission energies around $1.3$~eV. The QDs are embedded
in a field-effect structure to allow deterministic control of the
charge occupation of the dot \cite{Drexler1994}. On the `back'
side, a $25$~nm thick GaAs tunnelling barrier separates the QDs
from a heavily $n^+$ doped GaAs layer (thickness 20 nm, doping
concentration $4\cdot 10^{18}~\text{cm}^{-3}$) which forms the
back electrode. The `top' side of the QD-layer is covered first by
a $30$~nm thick GaAs capping layer, and then with an additional
AlGaAs/GaAs SL of $116$~nm thickness. A $5$~nm NiCr layer was
evaporated on top of the SL to form the second electrode.

\begin{figure}[!t]
\includegraphics[scale=1]{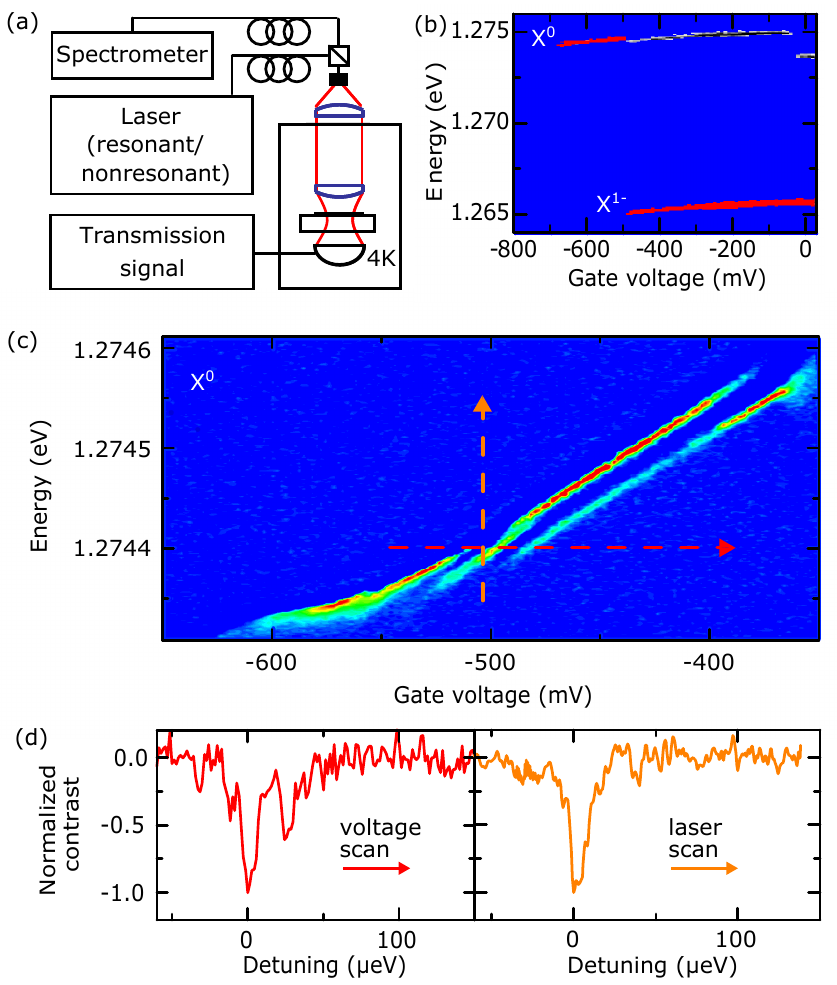}
\caption{(a) Setup for spectroscopy of a single quantum dot at
4.2~K. For resonant excitation, the transmission is measured by a
photodiode underneath the sample. Photoluminescence is measured
using a spectrometer. (b) Photoluminescence charging diagram of a
quantum dot (QD1) with characteristic X$^0$ and X$^{1-}$ stability
plateaus. (c) High resolution DT spectroscopy of the X$^0$
stability plateau for the same dot as in (b). In addition to the
linear Stark shift, there are several abrupt changes in the
exciton resonance energies as the gate voltage is varied. (d)
Normalized DT spectra along the dashed lines shown in (c).}
 \label{Fig1}
\end{figure}

The energy levels of individual QDs were investigated with
photoluminescence (PL) \cite{Warburton2000} and differential
transmission (DT) \cite{Alen2003} spectroscopy in a confocal
microscope setup shown schematically in Figure \ref{Fig1}(a). The
sample had a QD density such that there were typically $10-50$ QDs
in an area corresponding to a diffraction-limited focal spot of
$\sim 1 \upmu$m diameter. Out of this small ensemble, individual
QDs were spectrally selected for PL and DT measurements. By
applying a gate voltage \Vg~ between the top gate and the back
contact, the energy levels of a QD shift relative to the Fermi
energy E$_F$, allowing control over the number of electrons that
occupy the dot \cite{Drexler1994}. Figure~\ref{Fig1}(b) represents
a typical QD PL charging diagram as a function of \Vg, showing the
neutral exciton (X$^0$) and the negatively charged exciton
(X$^{1-}$) emission resonances, respectively \cite{Warburton2000}.

The neutral exciton transition was investigated in finer detail
for a number of QDs using DT spectroscopy, with the polarization
of the excitation laser chosen so as to excite just one of the two
exchange-interaction split resonances \cite{Hogele2004a}. Examples
of DT spectra are shown in Figure~\ref{Fig1}(d). The calculated
lifetime-limited linewidth of the X$^0$ transition is $\sim
0.7$~\mueV, however the transition is further broadened due to
charge fluctuations in the solid state matrix surrounding the QD
\cite{Hogele2004a,Kuhlmann2013}. For charge fluctuations that
occur on a time scale much faster than the measurement integration
times (typically $\sim$1~s), the resulting jitter in the QD
resonance energy is observed as broadened linewidths in the range
$4 \pm 2$~\mueV~ \cite{Hogele2004a,Kuhlmann2013}.

For a significant fraction of QDs, the linear relationship between
the exciton resonance energy and the applied gate voltage is
interrupted by several distinct jumps. An example of this effect
is apparent in Figure~\ref{Fig1}(c). The energy dispersion
gradient is consistent across the X$^0$ transition plateau,
however there are discontinuities in energy observed at specific
values of \Vg. With increasing \Vg, the transition energy jumps to
lower values by an amount in the range of $7$ to $38$~\mueV. Such
spectral discontinuities could be caused by similar environmental
charge-fluctuations which give rise to exciton line-broadening.
However, the spectral jumps studied here in more detail occur at
specific gate voltages, and correspond to larger resonance-energy
shifts in the QD transition.  Recently, the work of Houel
\emph{et~al.} \cite{Houel2012a} attributed these spectral jumps to
discrete charging of potential-traps located at the interface to
the SL. Our analysis detailed below suggests the presence of
additional potential-traps in the surrounding GaAs matrix.

\begin{figure*}[!t]
 \includegraphics[scale=1]{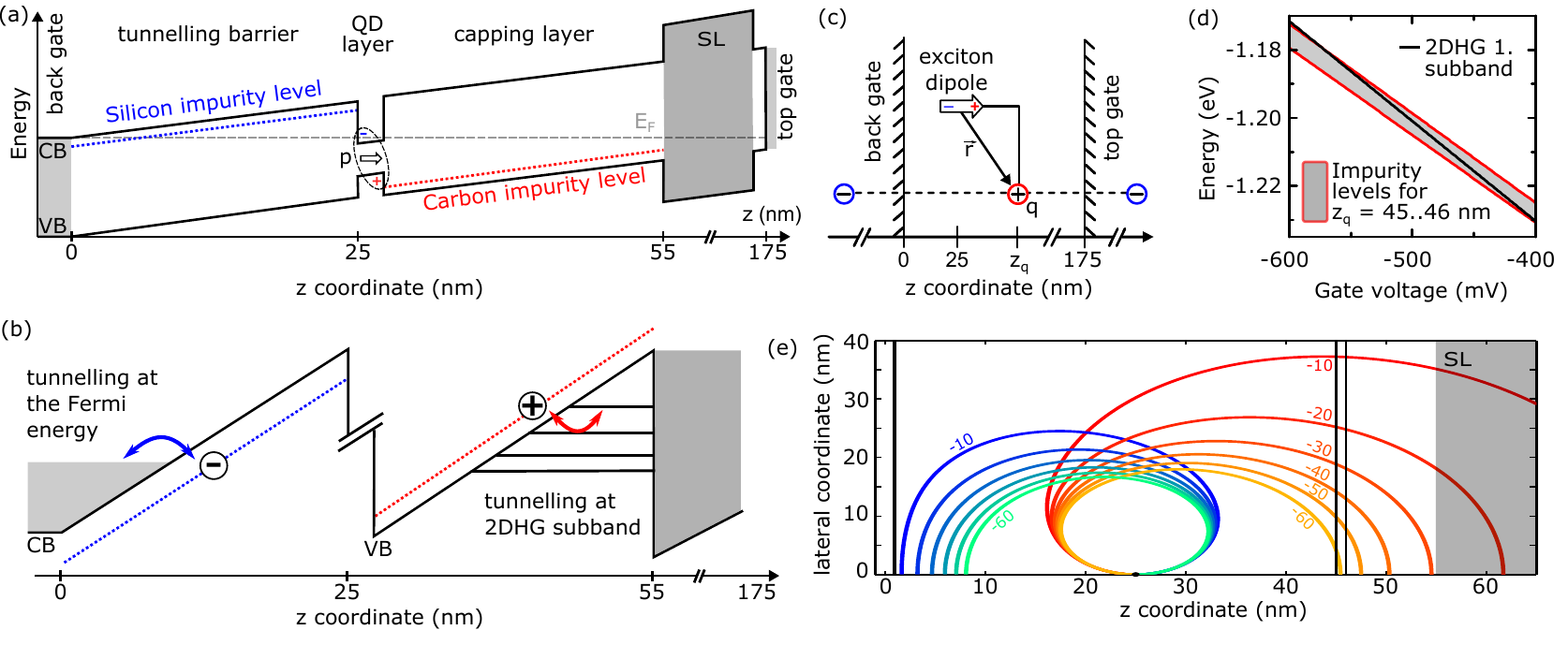}
 \caption{
(a) Conduction and valence band edges in the heterostructure
device. E$_F$ is the Fermi level, and SL labels the superlattice
region, which is scaled down for better visibility of the regions
of interest. The QDs are located at $z=25$~nm. (b) Impurity
charging processes that could give rise to the observed energy
jumps in the X$^0$ resonances; an electron-tunnelling resonance
between a silicon impurity and the back gate (left), or a
hole-tunnelling resonance between a carbon dopant and a subband in
the 2DHG that forms at the interface to the SL (right). (c) The
electrostatic model of the heterostructure. The exciton energy is
perturbed by an added charge $q$ (here shown to be positive).  The
effect of the conducting back- and top-gate layers are included to
first order in the form of image charges of opposite parity to $q$
(here negative). (d) The resonance condition at the valence band
edge between the carbon impurity level and the $n=1$ subband of
the 2DHG. The range of \Vg~ over which the resonance-jumps are
observed implies that the carbon sites must be located within the
$z_q=45-46$~nm region. (e) Positions of added charge $q$ that
result in specific values of energy shift $\Delta E$.  The contour
lines are labelled with the value of $\Delta E$ in \mueV, where
the orange-red lines are solutions for $q=+e$, and the green-blue
lines correspond to $q=-e$.  The grey lines of constant $z$
indicate the regions where there are tunnel resonances for either
the silicon or carbon impurity sites. It is important to note that
only the carbon impurities exhibit an overlap between the
tunnel-resonance \Vg~ range and the Stark-shift $\Delta E$ range,
and that this region of overlap is not at the interface to the
SL.} \label{Fig2}
\end{figure*}

\section{Modelling \& Discussion}

The exciton resonance energy $E$ of a QD is shifted by an electric
field $\textbf{F}$ through the quantum confined Stark effect:
\begin{equation}
E = E_0 - \textbf{p} \cdot \textbf{F} + \beta \cdot \textbf{F}^2
\label{eq:QCSE}
\end{equation}
where $E_0$ is the unperturbed exciton resonance energy,
$\textbf{p}=p \hspace{0.5mm} \hat{\textbf{z}}$ is the static
dipole moment of the exciton transition, and $\beta$ is the
polarizability \cite{Warburton2002a}. This relation quantifies how
an applied electric field can be used to shift the exciton
resonance deterministically, and also encompasses the mechanism
through which charge fluctuations in the solid-state matrix
surrounding the QD can perturb the resonance \cite{Houel2012a}.


An applied gate voltage $V_G$ generates an electric field
$\textbf{F}=\tfrac{-(V_G-V_S)}{l}\hat{\textbf{z}}$, where $l$ is
the distance between the $n^+$ layer and the top surface
electrode, and $V_S=0.62$~V is the Schottky barrier potential
\cite{Warburton2002a}. The axial polarizability $\beta_z$ along
the growth direction has been measured for similar dots as
$\approx -0.3$~\mueV /(kV/cm)$^2$ \cite{Warburton2002a}. For the
X$^0$ transition, which occurs with $|V_G+V_S|$ in the range
$0.9-1.1$~V, the effect of the axial polarizability is negligible
compared to the much larger dipole contribution. Therefore, the
magnitude of the dipole moment $p$ can be determined from the
gradient of the X$^0$ transition energy versus gate voltage $V_G$
as $p = (\partial E / \partial V_G)\cdot l$. Values of $p$ are
typically $e \times 0.2 \text{~nm}$ (where $e$ is the elementary charge) for the strongly confining QDs
surveyed in this work \cite{Warburton2002a}. In the QD plane,
there is no permanent dipole moment, however the larger geometric
extent of the dot in this direction implies a much larger lateral
polarizability of the order of $\beta_{xy} \approx -4$~\mueV
/(kV/cm)$^2$ \cite{Gerardot2007}. Therefore, charge fluctuations
in the vicinity of a QD can perturb the exciton resonance by
coupling to the permanent dipole moment in the
$\hat{z}$-direction, or through the polarizability in the lateral
plane.

The magnitude of the exciton resonance-energy shift caused by a
single unit charge $q$ placed near the dot can be determined with
a simple electrostatic model of the heterostructure depicted in
Figure~\ref{Fig2}(c). A QD exciton is represented as a dipole
oriented along the $\hat{z}$-axis, positioned between two
electrodes. A charge-trapping site is located at an arbitrary
distance from the dot, described by position vector $\textbf{r}$.
Upon occupation of such a trapping potential, the change in the
static electric field, $\Delta \textbf{F}$, at the QD position is
approximated as:
\begin{equation}
    \Delta \textbf{F} = \frac{1}{4 \pi \epsilon_0 \epsilon_r}
    \left(
    \frac{q}{|\textbf{r}|^2} \ \hat{\textbf{r}}
    +\frac{-q}{|\textbf{r}_{m_1}|^2} \ \hat{\textbf{r}}_{m_1}
    +\frac{-q}{|\textbf{r}_{m_2}|^2} \ \hat{\textbf{r}}_{m_2}
    \right),
    \label{eq:CoulombsLaw}
\end{equation}
where $q$ is a unit charge equal to either $\pm e$, $\epsilon_0$
is the permittivity, $\epsilon_r$ is the dielectric constant of
the surrounding GaAs matrix, and
$\hat{\textbf{r}}=\textbf{r}/|\textbf{r}|$. The first term in the
brackets arises from the impurity charge $q$. The response of the
freely-moving charges in the electrodes to the altered charge
environment is included (to first order) as second and third terms
in the form of image charges 
${m_1}$ and
${m_2}$, located behind the back gate and top electrode at
${\textbf{r}}_{m_1}$ and ${\textbf{r}}_{m_2}$ respectively [see
Figure~\ref{Fig2}(c)].

Combining equations~\ref{eq:QCSE} and \ref{eq:CoulombsLaw}, the
energy shift is obtained as a function of the position and parity
of the added charge. As an example, Figure~\ref{Fig2}(e) shows the
possible positions for added charges of either $\pm e$ that would
induce a step-change in exciton energy in the range of $-10$ to
$-60$~\mueV, calculated for the QD1 in Figure~\ref{Fig1} with $p=
e \times 0.208$~nm.
At large axial distances, the energy jumps could be caused by
either a negative charge appearing below the dot, or a positive
charge appearing above the dot, i.e. the observed charging events
produce an electric field which opposes the externally-controlled
field. In the lateral plane, the addition of either parity charge
could induce such an energy shift.
Aside from the magnitude
of the energy jumps, their gate voltage dependence is also central
to identifying the charge impurity location. Since the spectral
jumps occur at specific gate voltages, this suggests that the
individual trap sites are tuned through tunnel-resonances with
charge reservoirs as $V_G$ is varied.

On the lower side of the dot, the most likely source of the
electron-trapping sites are the silicon (Si) donor dopants. The
$n^+$ back-gate consists of heavily Si-doped GaAs, and previous
studies have shown that Si atoms diffuse during the growth process
up to several tens of nm along the growth direction of the sample
\cite{Kleemans2010}. The energy level associated with the Si
donor-electron lies $E_{\textrm{Si}}=5.8$~meV below the GaAs
conduction band edge \cite{Karasyuk1994}, and the possible \Vg-controlled
tunneling mechanism is a resonance with the Fermi level in the
back gate [see Figure~\ref{Fig2}(b)]. A Si impurity site with $z$
in the range $0.85$ to $1.0$~nm would be consistent with the
observed energy jumps occurring within the gate-voltage range of
$-600$ to $-400$~mV. However, a change of $-e$ at this location
would induce a QD resonance-energy shift less than $7$~\mueV ~[see
Figure~\ref{Fig2}(c)]. Such an energy shift is barely resolvable
within the X$^0$ linewidth, and indeed all the observed
discontinuities investigated here have a larger change in energy.
Therefore, we can exclude Si impurities as the origin for the
observed spectral jumps.

In the region above the QD layer, carbon (C) atoms are the likely
source of hole-trapping sites. There is inevitably a residual
background C-doping in any MBE grown device, and the concentration
is known to be on the order of $10^{15}$ cm$^{-3}$ for our sample.
The C acceptor atoms have an energy level $E_\text{C}=26$ ~meV
above the valence band \cite{Oelgart1990}. The \Vg-controlled
tunneling resonance in this case involves a sub-band in the
two-dimensional hole gas (2DHG) that forms at the interface to the
AlGaAs/GaAs SL [depicted in Figure~\ref{Fig2}(b)]. The energy of
the $n^{\text{th}}$-subband in the 2DHG is given by
\cite{Davies1998}:
\begin{equation}
    E_{\text{hole}}^{n}=E_{\text{gap}}+E_{\text{2DHG}}-c_n
    \left[\frac{(e\hbar)^2}{2m}F^2\right]^{1/3},
    \label{eq:E2DHG}
\end{equation}
where $E_{\text{gap}}$ denotes the GaAs bandgap energy,
$E_{\text{2DHG}}=e(V_G-V_S)/z_{\text{2DHG}}$ is the valence band
energy at the position of the 2DHG, $c_n$ is the $n^{\text{th}}$
Airy coefficient approximated by $c_n\approx\left[\frac{3}{2} \pi
(n-\frac{1}{4})\right]^{\frac{2}{3}}$, and the effective mass $m=0.57$ \cite{Bouarissa1999}.
The carbon charge-trap
energy as well as $E_{\text{hole}}^{1}$ as a function of \Vg~ are
shown in Figure~\ref{Fig2}(d), identifying resonance conditions in
the \Vg~ range from $-600$ to $-400$~mV for a carbon atom with $z$
in the interval of $45 - 46$~nm. These $z$-boundaries are depicted
in Figure~\ref{Fig2}(e) to highlight the fact that a charge of
$+e$ located within this $z$-slice can indeed induce energy shifts
up to -60~\mueV. This location of the charge traps is well within
in the GaAs capping layer and does not coincide with the interface
to the SL \cite{Houel2012a}. Remarkably, however, our results are
consistent with the observation that an increase of the separation
between the QD layer and the SL is sufficient to inhibit spectral
jumps in the plateau of X$^0$ and favor the narrowing of the
exciton resonance \cite{Houel2012a}. The displacement of the SL to
larger values of $z$ implies a change in the resonance condition
between the C-impurity level and the lowest 2DHG sub-band through
$z_{\text{2DHG}}$ in Equation~\ref{eq:E2DHG} such that
carbon impurity sites would effectively be depopulated at gate voltages
characteristic of the X$^0$ stability regime.

\subsection{Impurity-site charging dynamics}

\begin{figure}[!t]
\includegraphics[scale=1]{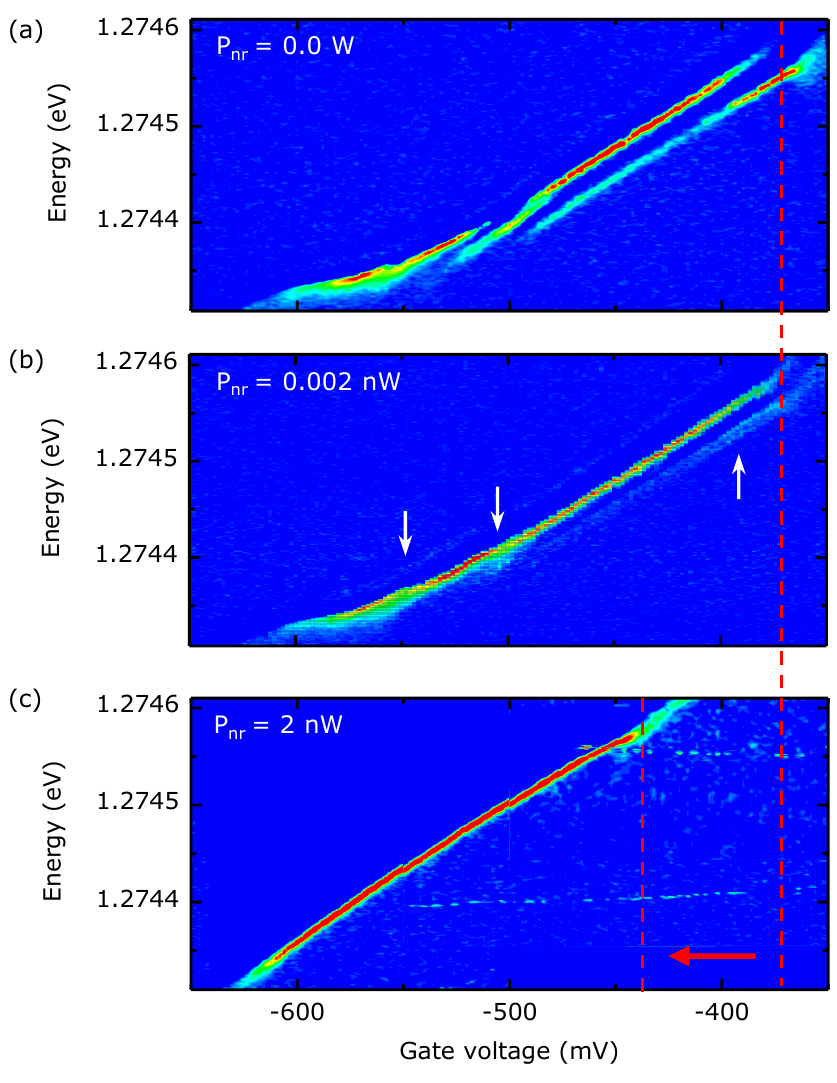}
\caption{X$^0$ stability plateaus of the QD1 recorded in DT with
an additional non-resonant laser at (a) zero, (b) low
($P_{nr}=0.002$~nW) and (c) high ($P_{nr}=2$~nW) power. (b) The
non-resonant laser at $850$~nm photo-generates charge carriers in
the wetting layer which yield reduced spectral jumps (indicated by
white arrows) due to partial saturation of the charge-trap
impurities. (c) At high non-resonant laser powers the charge
impurities are fully saturated and the plateau is free of spectral
discontinuities. Additionally, the accumulation of photoinduced
holes at the superlattice results in a shift of the stability
plateau (indicated by the red arrow) due to a partial screening of
the gate voltage.} \label{Fig3}
\end{figure}

The spectra of the X$^0$ transition for the dots surveyed in the
course of this work exhibited in general a more complex structure
than would be expected from the simplistic model described above.
The model explains the majority of spectral jumps, where the QD
transition energy changes abruptly from one value to a lower one
within a \Vg~ span of $5$ to $10$~mV. This overlap in gate voltage
of the QD energies corresponding to charged and uncharged impurity
states is indicative of the rate at which hole-tunnelling occurs
between the impurity site and the 2DHG. A fast tunnelling process
yields a small overlap in \Vg~ and vice versa, analogous to the
overlaps observed in PL charging diagrams of QDs in samples
with different thicknesses of the tunneling barrier between the QD
and the Fermi reservoir and correspondingly different electron
tunneling rates. In addition to the `sharp' transitions
in the charge state of the impurity site, we observe in our DT
data instances where an impurity-site coexists in both charged and
uncharged configurations over an extended gate voltage range of
$50$ to $100$~mV. An example of such a coexistence can be seen in
Figures~\ref{Fig1}(c) and \ref{Fig3}(a) for \Vg~ between $-480$~mV
and $-400$~mV. This behavior is inexplicable within the modelling
framework developed above. A refined model should take into account not only resonant tunneling between the impurity site and the 2DHG, but also dynamic charge capture processes that occur in the presence of an optically generated charge reservoir \cite{Nguyen2013}.



To qualitatively understand the impurity site charging dynamics,
we adopt the rate-equation formalism of Ref.~\cite{Nguyen2013} to
determine the time averaged steady-state occupation of the
impurity site $N_i$ as:
\begin{equation}
    N_i = \frac{1}{1+\gamma_e/\gamma_c},
    \label{eq:RateEqnNi}
\end{equation}
where $\gamma_c$ and $\gamma_e$ denote the rates at which a hole
is captured in, or escapes from, the impurity trap respectively.
In the simple case that was modelled above, $\gamma_e \gg
\gamma_c$ when the impurity site is energetically higher than the
$n=1$ subband of the 2DHG, and $N_i \rightarrow 0$. Conversely,
when the gate voltage is tuned such that the impurity site is
below the lowest 2DHG subband, then $\gamma_e \ll \gamma_c$, and
$N_i \rightarrow 1$.  However, the capture rate $\gamma_c$ can
also be influenced by the excitation of charge carriers in the QD.
Previous investigations have shown that the tunnelling rate of
holes from a QD is significantly enhanced as it is tuned through
resonances with energy levels in the 2DHG \cite{Seidl2005}.  It is
possible therefore, for holes to tunnel from the QD to an $n > 1$
level of the 2DHG, and then occupy the impurity-site before
finally relaxing to the energetically favorable $n=1$ state of the
2DHG.  These QD--DHG resonances effectively enhance $\gamma_c$
such that it becomes comparable to $\gamma_e$ (determined only by
the valence band properties), and therefore it becomes feasible
for the impurity-site to be partially occupied over an extended
\Vg ~range, despite not being resonant itself with the 2DHG state.
The QD--2DHG resonances observed in similar heterostructures were
measured to occur over a range of $\approx$100~mV in \Vg~
\cite{Seidl2005}, in agreement with the \Vg~span in which we
observe intermediate values of $N_i$.

We can further dynamically perturb the charge environment of the
system with the use of non-resonant optical excitation
\cite{Houel2012a}. In addition to the resonant laser, the output
of an 850~nm laser diode is directed onto the sample, which
excites electron-hole pairs in the wetting layer.  The effect of
this additional charge-carrier generation on the QD charge sensing
phenomena is two-fold, firstly altering the electrostatic response
of the QD, and secondly directly influencing $N_i$. The first of
these effects is due to a build up of holes in the 2DEG, which are
generated in the wetting layer, but due to the energy-gradient
across the heterostructure, tend to relax into the 2DHG.  This
accumulation of positive charge at the interface to the SL has the
effect of partially screening the dot from the externally applied
field (causing the well known energy shift of the exciton plateau
\cite{Smith2003}) as well as screening the QD from the impurity
charge.  Consequently the spectral jumps decrease monotonically in
magnitude with increased non-resonant laser power (see
Fig.~\ref{Fig3}). The effectiveness of the screening depends on
the charge density of the 2DHG, and therefore is determined by the
laser power. In the limit of high charge density, we can modify
the electrostatic model to include the response of the 2DHG in the
form of an additional mirror charge. With this modification, an
energy jump in the QD spectra of 30~\mueV~in the absence of
non-resonant light is reduced to just 16~\mueV. In addition to
this electrostatic shielding, the second effect of non-resonant
excitation is the direct influence on $N_i$ \cite{Nguyen2013}.
The capture rate $\gamma_c$ is increased with $P_\text{pump}$, and
beyond a certain saturation power $N_i \rightarrow 1$, despite the
impurity trap not being energetically favorable compared to the
$n=1$ level of the 2DHG.  This dynamic saturation effect can be
observed in Figure~\ref{Fig3}(c).

\subsection{Single impurity sensed with multiple QDs}

\begin{figure}[t]
\begin{center}
\includegraphics[scale=0.85]{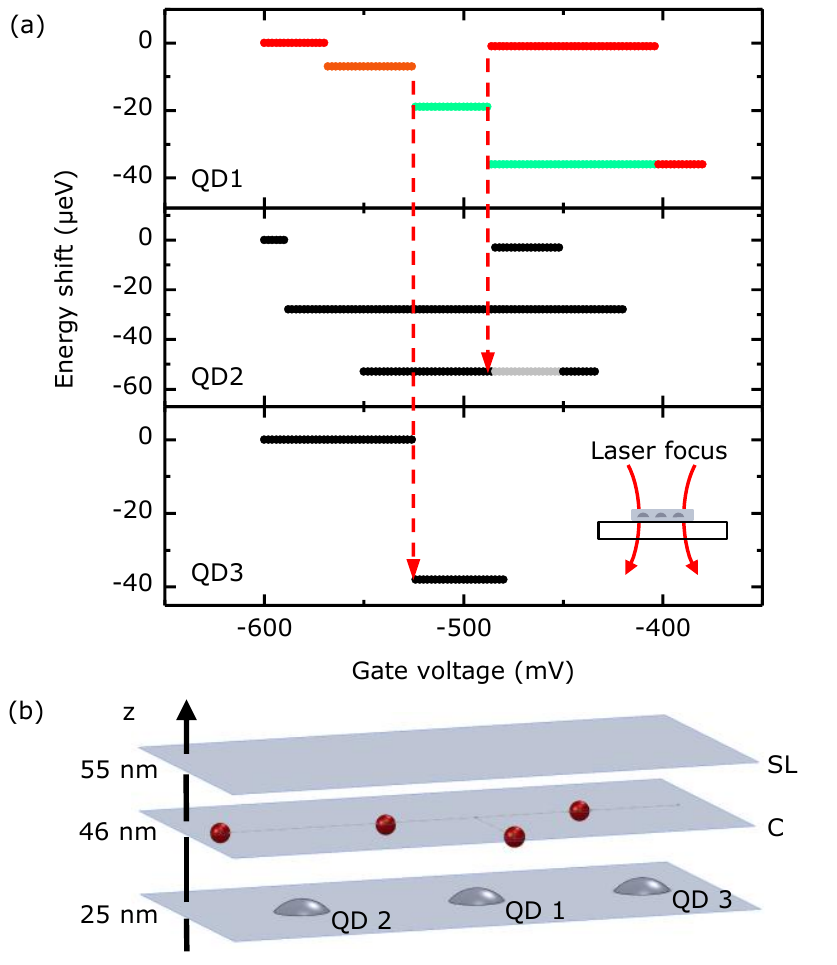}
\caption{(a) Plateau of the neutral exciton for three different
quantum dots. The linear Stark shift is subtracted for clarity.
All three quantum dots are situated within one common laser spot
(inset lowest panel). QD 1 shows three distinct jumps in the
exciton dispersion, QD2 has two and QD3 features one
discontinuity. QD1 and QD2 have one jump in common which is at a
gate voltage of -485 mV and for QD1 and QD3 a joint discontinuity
at -526 mV is observed. (b) 3D-model of the three quantum dots
with corresponding impurities assuming that only carbon states
participate. This is only one possible configuration as the angles
$\varphi$ between the quantum dots and impurity cannot be
determined. Impurities are red and the quantum dots are grey.
}\label{Fig4}
\end{center}
\end{figure}

The transverse location of an impurity site cannot be pinpointed
with just one QD sensor, however further constraints can be
obtained by using multiple QDs within the impurity vicinity. The
absorption spectra of three different QDs (labelled as QD1 to QD3)
within a common focal spot (of $\approx$1~$\upmu$m in diameter)
are shown in Figure \ref{Fig4}(a), with the background linear
Stark shift subtracted for clarity.  It can be seen that there are
concurrent spectral jumps occurring for two different QDs at the
same gate voltage \Vg, which are likely caused by one single
impurity.  For the spectral-jumps which only occur in a single
spectrum we assume the charge-trapping sites are too far away from
the alternate dots for the spectral effects to be resolved.

As a specific example of the charge sensing capability, we
determine the location of the impurity site that causes jumps in
the spectra of QD1 and QD3.  The $z$-position of the charge
impurity is calculated (using \Vg$=528.5 \pm 3$~mV) as $z_q=46.0
\pm 0.5$~nm.  The magnitude of the energy-jump in the QD1 spectrum
is $\Delta E = 12 \pm 2$~\mueV, while for QD2 the energy change is
$\Delta E = 38 \pm 2$~\mueV.  The measured dipole moments for each
of these dots ($p= e \times 0.180 \text{~nm}$ and $e \times 0.208
\text{~nm}$ for QD1 and QD3, respectively) determine the
transverse location of the impurity site as $29.4 \pm 6.3$~nm from
QD1, and $12.2 \pm 3.0$~nm from QD3.  If the relative locations of
the QDs were known, there would be a unique solution for the
location of the charge impurity.  This concurrent sensing concept
is depicted in Figure \ref{Fig4}(b), showing the QDs linearly
aligned, and a number of carbon impurity sites in the GaAs volume
between the QD layer and the SL.

\section{Conclusion}
In summary, we have identified the cause of spectral jumps in the
neutral exciton transitions of QDs as being due to charging of
carbon impurity sites. Our results suggest these impurities are
located in the semiconductor region surrounding the QD layer. This
is further re-enforced by measuring the spectral signatures of the
charge trapping concurrently for more than one QD. Despite the
fact that the charge trapping sites are not themselves at the
interface, our analysis suggests that moving the SL interface
further from the dot layer would still improve the exciton
resonance stability, by shifting the tunnel resonances to
different gate voltages.

\section{Acknowledgments}
We acknowledge funding by the Deutsche Forschungsgemeinschaft
(SFB~631 and the German Excellence Initiative via the Nanosystems
Initiative Munich, NIM) and support from the Center for
NanoScience (CeNS).


\end{document}